# Quality of Life and the Experience of Context

Ankur Betageri
Bharati College, University of Delhi,
New Delhi, India
0000-0002-8737-7542

# Abstract

I propose that quality of life can be compared despite the difference in values across cultures when it is experienced at the sensory and perceptual level. I argue that an approach to assessing quality of life which focuses on an individual's ability to organize his or her context by perceiving positive constellations of factors in the environment and his or her ability to achieve valuable acts and realize valuable states of being is more meaningful than the approaches of metrics which focus directly, and often solely, on the means of living and the means of freedom. Because the felt experience of quality of life is derived from a constellation of factors which make up the indivisible structure of a milieu, the experience of quality of life cannot be regarded as a subjective experience. Through the example of how different frequencies, and mixtures of frequencies, of light are perceived as colour by the eye, I demonstrate that the human cognitive apparatus, because of its relation to the object that is measured, apprehends different scales of quantity as degrees of quality. I show that lived experience is the result of a selective relationality with one's environment and that the experience of quality has something to do with the perception of entities in their interrelated and networked nature as wholes.

**Keywords:** Perception of quality, Organizing context, Quality of life, Sensory translation, Selective relationality.

# 1. Introduction

Most developed societies offer the possibility of fitting into alternative traditions. We can at least recognize the binary of traditional/alternative approaches in most developed cultures: a traditional approach is guided by the sedimentation of forms of life over centuries, and an alternative relational approach is formed in opposition to the traditional approach by considering, to use a phrase by Gregory Bateson, the organism plus the environment as the unit of survival. It is therefore important to know what we think of when we think of the environment or how our mind organizes a frame of reference, or a context, for itself. When we use the word context we use it to refer to a "location outside the physical individual." (Bateson 1987: 256). But there is another kind of 'context,' the power of organizing a frame of reference, which lies not outside the physical individual but is rather inside him: this power of organizing a frame of reference is derived from orders of learning. Organizing a frame of reference which involves construction of sign-systems that can accommodate perceptual experience and connect it with already existing sign-systems requires a higher order of learning. But belonging to a frame of reference requires learning from others. Learning from others is the most important mode of adaptation among human beings; the sheer scale of this form of adaptation makes human beings unique among animals.

Uexküll proposed a similar idea but expressed it in biological terms: "All animal subjects, from the simplest to the most complex, are inserted into their environments to the same degree of perfection. The simple animal has a simple environment; the multiform animal has an environment just as richly articulated as it is." (Uexküll 2010: 50). For Uexküll, a subject's environment is "the closed unit formed by two worlds: its perception world [*Merkwelt*] comprising of everything the subject perceives and its effect world [*Wirkwelt*] made up of everything it produces." (Uexküll 2010: 42). Uexküll imagined animal and human subjects as being enclosed in individual soap bubbles made up of subjective perception signs and effect signs. No space, according to him, is independent of subjects. But these bubbles of subjective perception signs that individual subjects inhabit effortlessly overlap one another. (Uexküll 2010: 70). So, for an animal with a complex mind like the human, there are as many features in the environment as he is able to perceive, and attach signs to, and as many effects in the environment as he is able to produce.

At higher orders of scientific learning one organizes one's frame of reference by participating in the thought structure created by, what Ludwik Fleck calls, a 'thought collective' which subscribes to a 'style of thought'. Style of thought refers to a "particular way of thinking, seeing, and practicing" which "involves formulating statements that are only possible and intelligible within that way of thinking." (Rose 2007: 12). In a style of thought, arguments and explanations are elaborated through concepts, terms, references and relations to explain phenomena which are already classified and sorted in a certain way. Certain things are designated as evidence and they are gathered, measured and used in specified ways. Model systems are then created by assembling data and projecting them in the form of charts, tables, and graphs. This style of thought is connected to practical arrangements such as experiments and clinical trials to provide empirical validation. A style of thought belongs to a discipline or subdiscipline and involves membership of a 'thought community.' (Rose 2007: 12). Fleck's idea of thought community can be compared to Peirce's idea of *commens* which refers to the common mind into which the minds of the utterer and interpreter have to be fused for any communication to take place. *Commens* "consists of all that is, and must be, well understood between utterer and interpreter, at the outset," for the sign to "fulfill its function." (Peirce 1998: 478).

The concept of affordances proposed by James Gibson is useful in understanding the relational situation in an environment. Affordances are properties of an environment which allow an organism to use its abilities to take advantage of the resources that the environment holds. They are the conditions in an environment which afford functionality and support the abilities of an organism; the co-occurrence of supportive properties in an environment and abilities in an organism ensures the optimum functionality of the organism. Higher orders of learning in humans allow them to recognize the affordances – or their lack – in an environment as it is given. When an individual with a higher order of learning recognizes the lack of affordances in an environment as it is given he begins to use the power to organize a frame of reference for himself.

## 2. The Capability Approach to Quality of Life

The most difficult thing to establish in the contemporary world is the idea of a universal standard to measure the quality of life which is independent of statistical standards founded on traditional socio-economic criteria. There are many statistical standards that national governments use to indicate development and wellbeing like gross domestic product (GDP) which "mainly measures market production – expressed in money units" but which "has often been treated as if it were a measure of economic well-being" (Stiglitz et al. 2010: 11), and its alternative, the Human Development Index (HDI) which measures the capability to live a long and healthy life and achieve a basic minimum standard of living. To identify the limits of GDP as an indicator of economic performance and social progress The Commission on the Measurement of Economic Performance and Social Progress (CMEPSP) was set up in 2008. The unifying theme of the report submitted by the Commission was that "the time is ripe for our measurement system to *shift emphasis from measuring economic production to measuring people's well-being.*" (Stiglitz et al. 2010: 10)

What can be posited in opposition to statistical standards of living is the concept of *quality of life* which I understand as the state of well-being that a person experiences by selectively relating to the environment. A measure that focuses on quality of life relies on felt experience; it measures quality of life in terms of a person's evolvability in perceiving positive constellations of significance in the environment through the acquisition of knowledge and skills and through the exercise of his capability. Quality of life is determined by a person's ability to perceive the constellation of factors in a milieu that afford his capabilities and sensibilities and by his ability to achieve valuable acts and realize valuable states of being. Metrics that measure standard of living are statistical standards which involve the direct use of quantitative data and statistical measures to measure the material standard of life. They attach direct importance to the *means* of living and *means* of freedom to the well-being of the individual quantifying them in terms of real income, wealth, primary commodities and resources. But if we focus on a person's capability— defined broadly as a person's freedom to choose what he wants to do and be—Sen argues, we assess the quality of life directly in terms of an individual's capacity to achieve valuable acts and realize valuable states of being, and value these acts and states as central to living, attaching only a derivative importance to the means of living and the means of freedom. (Sen 1993: 30-33).

An important aspect of a person's capability is his communicative freedom that is capable of conveying the materiality and intensity of experiential reality, that is, in other words, any kind of symbolic freedom that is capable of virtualizing the actuality of experience. The capacity to communicate information in a way that it reflects the interrelated and networked nature of reality and which enables a meaningful and immersive experience plays a very important role in a person's realizing valuable acts and valuable states of being.

Quality of life which is understood in terms of life satisfaction and which is based on felt experience is derived from a constellation of factors that make up an ecological system. These constellations of factors are not given in an environment but are chosen by the individual subject because they afford his capabilities and interests. The capabilities of a person are something apart from what Sen calls functionings; functionings are aspects of the state of a person, the different things that a person manages to be, like being well-fed and in good health, and the various things he manages to do—while being employed and to be socially integrated—in leading a life. Capabilities, on the other hand, constitute an important aspect of a person which extend the scope of individual initiative and freedom. Quality of life is determined by a person's ability to organize a context that affords his capabilities and in which he achieves valuable acts and realizes valuable states of being. The world-richness of the curious and the imaginative who have the capacity to achieve valuable acts and realize valuable states of being cannot easily be quantified.

One can measure quality of life by mapping the constellation of factors which make up a milieu in terms of felt experience and by assessing to what extent they are valuable to an individual in realizing valuable acts. The life satisfaction derived from living in a certain geographical area is different from that derived from living in a different geographical area: what differentiate two different geographical areas in the context of quality of life are the constellations of factors that support the capacities of a person. The constellation of factors need not only be aspects outside the individual, like the presence of social conditions necessary for thriving, but may include substances and parts of diet that affect a person at the molecular level and bring about a positive change in his mood, emotions, and behaviour.

The state of health requires regularity in the functioning of the organs in the body. When these organs in the *milieu intérieur* or internal environment are provided with the required nutrients their performance is enhanced leading to the all-round betterment in the state of general health and an increased efficiency in the performance of the chosen task. The internal health of a person expresses itself in terms of a person harmonizing with the external environment by selectively relating to elements in the environment that afford his thought and behaviour. The state of internal health is thus one that orients a person to realize a chosen task or experience a valued state of being.

Surveys that monitor social conditions like Eurobarometer and Human Development Index (HDI) measure quality of life, but they are nationalistic standards; quality of life as a measure can be made meaningful and closer to lived experience by making it a regional standard, the region being defined by the constellation of features that make up the milieu of an individual. People who straddle two or more different life-worlds are in a better place to compare life-satisfaction; their sense of quality of life is spontaneous and immediate and is founded on lived experience. So, by measuring life satisfaction on a constellation of indicators that make up the experience of living in a milieu we can arrive at the measure of quality of life.

# 3. A Gestalt Approach to Arrive at the Concept of 'Quality of Life'

Analogous to urban life is life in the countryside which offers a completely different mode of life as well as quality of life. The life of many children who grow up in urban localities is enriched by visits to the countryside, especially if they have an ancestral family living there. So regardless of the way countries are classified, as developed, and developing, children (as opposed to adults who have a specific role to play in the socio-economic system) living in these countries have analogous experiences. Life in urban centers as opposed to life in communes in France, cabin life as part of the classic *friluftsliv* (*lit*. 'free air life'; 'open air life') versus living with 'ordinary' ways of using energy in Nordic countries are analogous – though not in an economic, and perhaps even 'material' sense – to the lived experience of individuals from developing countries like India. This means the concept of *quality of life* derived from lived experience in a region is a more universal measure than the concept of the standard of living derived from inflation adjusted income and other factors such as access to healthcare, availability of employment, and life expectancy. What we experience as the quality of life is dependent on a relational field where any concrete content of experience can only be measured by relating it to an "indivisible structure, a *constellation* of factors." (Naess 2001: 57). And because what we experience as reality is constituted by these concrete contents and abstract structures our felt experience of the quality of life cannot be regarded as a subjective experience. The world of concrete contents which gives us the experience of the quality of life therefore has a gestalt character rather than atomic character. It is by joining the dots of the constellation of factors that we get a frame of reference, and there are no better frames of reference, says Naess, than those provided by gestalts. (Naess 2008: 80).

## 4. The Appearance of the Path

The path marked by a constellation of factors in the human environment whether planned by nature or one recognized in natural and social circumstances may present itself as a magical formation to the individual subject. In the preceding sections I have focused on the recognition of the constellation of factors in natural and social formations that makes a path visible to the individual subject. Uexküll calls the path planned by nature the inborn path. The inborn path is the path that has been laid out by nature and which occurs at a particular stage in the lifespan of a subject but occurs as a magical formation from the point of view of the individual subject. The path that the young pea weevil (*Bruchus pisorum*) sees out of the hardened pea pod into whose soft flesh it had bored a channel in its pre-metamorphosis stage as a larva is an example of the inborn path. For the young pea weevil the way out does not occur as an inborn path but stretches out clearly marked as a magical formation. (Uexküll 2010: 122).

The curvy line that a female birch-roller (*Deporaus betulae*) cuts into a birch-leaf, which she later rolls into a sac into which she will lay her eggs, is another example of an inborn path. Though the curvy line which needs to be cut to roll the birch-leaf into home has never been cut by the beetle before – there being no indication of such a line on the leaf – the path that needs to be cut presents itself magically to the individual beetle.

The flight path of migratory birds that is borne by continents but which is visible only to the birds is another example of an inborn path. To a young flock of birds that have never migrated before and that have to make their way across continents without help from their parents, the flight path occurs as a magical formation. Birds are said to navigate and find their migrational flight path by following the intensities of earth's magnetic field. (Wiltschko and Wiltschko 2019: 1) But that birds navigate because of their capacity for geomagnetic reception or because they experience circannual biological rhythms in their bodies which make them adopt migrational behaviour (Gwinner 1986: 1-8) does not mean that Uexküll's explanation is wrong: the birds may still *experience* the long migrational flight path, which they discover through their first flight, as a magical formation.

The inborn path is different from what Uexküll calls the familiar path. The familiar path is a path traversed by the subject through the recognition of perception signs and effect signs established through previous experiences that follow one after another. The inborn path, on the other hand, is one in which a series of signs is immediately given as a magical phenomenon. The main difference between the familiar path and the inborn path is that while the outwardly transposed perception signs and effect signs of the familiar path are activated by sensory stimuli, the inborn path sounds out one after another like an inborn melody. (Uexküll 2010: 124).

## 5. Colour Sensation as the Sensory Translation of Quantity as Quality

Colour sensation is produced by the electromagnetic radiation of light when it is between the wavelengths of 800 to 400 millimicrons. When the wavelength of light is more than 800 millimicrons we get the infrared, and when it is less than 400 millimicrons we get the ultraviolet—both of which are invisible to the eye. (Schrödinger 2017: 153).

Colour sensation, then, is a sensual quality that is produced by electromagnetic waves of light in relation to the apprehensive mechanism of the human eye. The mixture of different wavelengths of electromagnetic radiation (expressed in terms of quantity) produces different colour sensations which constitute the immediate and spontaneous apprehension of quantity as quality. The mixture of different wave-lengths of electromagnetic radiation produces colours which are indistinguishable from the ones produced by the wavelengths of a single spectral light. The mixing of the wavelengths of radiation of red and blue from the extremities of the colour spectrum also gives us the colour purple which is not produced by any single spectral light. (Schrödinger 2017: 154).

According to Newton, the pure colours of the visible spectrum, revealed when a beam of light is passed through a prism, were elementary components of light that added up to produce white light. But he soon realized that the colours corresponded to frequencies. If light is regarded as independent of the human visual perception, there are only different frequencies of electromagnetic radiation of light; it is the human eye and the optic centre in the brain which make colour sensation possible. It is the eye which translates the different frequencies of light, and the mixture of different frequencies of light, as colour. And colour is a quality.

## 6. Lived Quality of Life is Perceptual and Relational

Because of the relation of the human apprehensive mechanism to an object that is measured there occurs a transformation of the measured object from a scale of quantity to a degree of quality. This transformation happens when we suspend the economic attitude (which informs our understanding of the standard of living) and adopt the aesthetic and ecological attitude. The aesthetic attitude, as Bateson (1979: 8) says, is that which is responsive to the pattern which connects. Quality is something that emerges out of our perception and appreciation of the pattern that connects parts to a whole like shapes, forms and relations and forms a whole. Degrees of quality are primarily apprehended as sensual and intellectual experiences and are expressed in terms of degrees like high, moderate, and low. Unlike quantity, the idea of quality is relational: it is something that expresses itself in terms of perceptual units. Because when we measure something, we measure it not only in terms of, but in relation to, something, and the something in relation to which standard of life is measured is the human. And the value of the human in an environment (on which his or her standard of life depends) is only ascertained in relational terms; it is only then that the value is supra-subjective in principle and inter-subjective in fact. Yet, when standard of living is calculated, it is calculated according to quantitative socio-economic parameters, and in the process what it leaves out is the relational perceptual dimension—which the human cognitive system and sensual system have a way of apprehending. This means when we take into consideration the relational aspect, the satisfaction derived from living in a certain system must be expressed in terms of quality of life. Measured quantitative data can express the status and dynamics of a socio-economic milieu but these numbers and statistical data, in the context of life satisfaction, are experienced as qualities, and these felt qualities do not have a one-to-one correlation with standard of living which is calculated on the basis of quantitative data related to income, housing, energy consumption, access to healthcare, life expectancy etc. In short, felt quality of life is dependent on the constellation of factors that make up the reality of a given milieu and this dependence is often expressed as the positive correlation between subjective well-being and objective conditions of living but standard of living which is calculated on the basis of quantitative data on socio-economic parameters, because of its nature as a 'standard', tells us not so much about the *lived quality of life* as it does about the reality of a nation and its population at the level of abstraction.

When we say a careful perception of external environment heightens the quality of life what we mean is the individuality and singularity granted to the combination of elements in the external environment draws an affective response that is equal in intensity to the singularity of the external being. The affective response is a sensory and organic response of the organism that is generative and productive of sensory organs. Just as the eye was evolutionarily shaped by the presence of light the presence of a singularity of the combination of elements in the environment shapes perceptual modes that organize the content and intensity of our Umwelten.

# 7. Affective States and the Appreciation of Becoming

Felt experience and valuable states of being are affective states or intensities and are processes of intensification which involve the crossing of intensive thresholds with the body. The experience of intensification of affect is not the experience of sensible qualities but the experience of intensity itself which involves transitions that reflect gradations of intensity. At the level of the body one can understand the transitions in intensity as variations in metabolism, in the speed of thought, and in the experience of affective sensory intensities—intensities contained in qualities like sound, music, taste, colour, temperature etc.; in signs, symbols and narratives and in the singularities that shape personalities, events and temporalities.

Intensification is the outcome of capacities to affect and capacities to be affected that characterize each thing in a given ecosystem. Affective states can be understood in terms of relations of speed and slowness of perceptions, variations in metabolism, actions and reactions of bodies, animals and humans. Since affective states or intensities are, as Deleuze calls them, "an implicated, enveloped or 'embryonized' quantity" (Deleuze 1994: 237) one cannot add them up or divide them without changing their nature. Intensities such as temperature, colour, tastes, light, pressure, speed etc., and intensities which are enveloped or implicated in a sign and which give birth to affective states, through their capacity to affect us, and our capacity to be affected by them, are intensive qualities that cannot be counted and divided without changing their nature and the affective states that they give birth to. Also, affective states in individuals belong to a dimension which is immanent to a space where space itself is a field of heterogeneous intensities; this is different from the experience of extensive quantities. Quality of life understood in terms of realizing affective states or intensities that are valuable in themselves involve the crossing of intensive thresholds that can only be appreciated in relation to the human, by employing an aesthetic and ecological attitude, and are therefore understood in terms of degrees of quality.

The experience of intensity of affect is usually associated with the loss of conventional coordinates and with the loss of self but this loss of conventional coordinates is followed by the formation of novel coordinates and the conceptualization of a dynamic and expansive self. The experience of a dynamic and expansive self as a becoming-X, as the flow of decoded desire that affects and gets affected by different entities in the environment, ends in the achievement of creative acts and in the realization of intense states of being.

## 8. The Perception of Quality

There is a universal dimension to the idea of good life; the universality of the basic notion of the good life comprises the three important determinants of state of health, income, and family life (Noll 2018: 10). But the universalism of lived experience itself can only be accessed through the ecological perspective. When lived experience is analyzed and decomposed by imposing the schema of cause-effect relationship, as an *effect* caused by empirical parameters or by cultural specificities, we not only lose the sensual pleasure which connects us through shared experience, we lose the very character and meaning of lived experience. Our perception of quality is not the result of a cause-and-effect relationship; there is no such thing as the automatic generation of quality from a cause which is independent of perception. Quality is experienced through a mode of relating to the external environment, that is, by aligning one's nature and behaviour to selectively apprehend certain aspects in one's environment. It is this selective relationality with one's environment which creates the lived experience. The existence of a certain measurable and quantifiable empirical reality, and the identification of cultural attitudes and the general schemas of behaviour of a population through which the empirical reality is filtered, does not give any idea about the individual experience of quality. Quality of life is something that is selectively apprehended and lived from the generality of experience of the populace which is assumed and posited to confirm the specificity, singularity, and value of one's own experience.

Perception of interrelationships in nature happens at the perceptual level by positing representations, that is, metaphors and symbols, which have a vitality and aliveness of their own. This requires the reconstruction of original perception and a delayed decoding of sense data in accordance with one's thinking style, a literary form, one's knowledge of the state of affairs and a cultural ethos. Symbols, and images generated by singular events, act as semantic objects that demand interpretation—these semantic objects, interacting with the human imagination often give birth to myths. What human imagination, individual and collective, brings to perception is this metaphorical dimension which we can see most vigorously expressed in mythology and folk imagination. Myths are extra-territorial environments, and mythical beings are extra-territorial forms, that can seed territories and make them look familiar and 'uniform'. What metaphorical perception conceals, which the scientific attitude pries and examines, is the network of relationships and the interrelatedness between entities which make them living wholes.

If we still resemble our mythologized and extinct ancestors it is because we occupy ecological and cultural niches similar to their own. And as our niches transform and develop, our interpretation of these myths and mythological ancestors also change and develop. Populations that occupy an ecological and cultural niche have a tendency to maintain the identity of the niche with the use of sign-systems and narratives. This tendency of populations isolated by geography and language to maintain their own distinctive identity is analogous to the biological process of speciation. This explains the persistence, as well as transformation, of communities – provincial, linguistic, ethnic, nationalistic – and cultural groups across generations.

Environments, whether natural or built, are capable of being perceived in two ways: as metaphorical wholes through composition, or as discrete entities with specific relationships (causal, correlative etc.,) through de-composition. Anthropocentric thinking, which science regards as an error of perception, is something that is central to mythological/mythopoeic thought. The experience of quality too has something to do with the unconscious formation (and conscious composition) of metaphors and images so that entities are perceived in their interrelated and networked nature as wholes. If the occurrence of metaphors is the first step in the composition of quality and meaning, narrative is that which connects the metaphors and images to mirror and capture the living stream of experience.

**Conflict of Interest**
The author has no conflict of interest to declare.

Wiltschko, R and Wiltschko, W. 2019. Magnetoreception in Birds. *J. R. Soc. Interface* 16: 20190295. DOI: 10.1098/rsif.2019.0295